\documentclass[reprint,superscriptaddress]{revtex4-1}
\usepackage{amsmath,graphicx}

\begin{document}

\title{Optical Wireless Information Transfer with Nonlinear Micromechanical Resonators }
\date{28 August 2017}

\author{Joseph A. Boales}
\affiliation{Department of Physics, Boston University, 590 Commonwealth Avenue, Boston, MA 02215, US}
\author{Farrukh Mateen}
\affiliation{Department of Mechanical and Aerospace Engineering, Boston University, 110 Cummington Street, Boston, MA 02215, USA}
\author{Pritiraj Mohanty}
\affiliation{Department of Physics, Boston University, 590 Commonwealth Avenue, Boston, MA 02215, US}

\begin{abstract}
Wireless transfer of information is the basis of modern communication. It includes cellular, WiFi, Bluetooth and GPS systems, all of which use electromagnetic radio waves with frequencies ranging from typically 100 MHz to a few GHz. However, several long-standing challenges with standard radio-wave wireless transmission still exist, including keeping secure transmission of data from potential compromise. Here, we demonstrate wireless information transfer using a line-of-sight optical architecture with a micromechanical element. In this fundamentally new approach, a laser beam encoded with information impinges on a nonlinear micromechanical resonator located a distance from the laser. The force generated by the radiation pressure of the laser light on the nonlinear micromechanical resonator produces a sideband modulation signal, which carries the precise information encoded in the subtle changes in the radiation pressure. Using this, we demonstrate data and image transfer with one hundred percent fidelity with a single 96 micron by 270 micron silicon resonator element in an optical frequency band. This mechanical approach relies only on the momentum of the incident photons and is therefore able to use any portion of the optical frequency band—a band that is 10,000 times wider than the radio frequency band. Our line-of-sight architecture using highly scalable micromechanical resonators offers new possibilities in wireless communication. Due to their small size, these resonators can be easily arrayed while maintaining a small form factor to provide redundancy and parallelism.
\end{abstract}
\maketitle

\section*{Introduction}
Micro- and Nano-Electro-Mechanical Systems (MEMS and NEMS) resonators are of fundamental and technological interest, with applications ranging from timing in integrated circuits \cite{nguyen2005} to methods in quantum metrology \cite{boales2016,mancini2003}, while operating in their linear regimes. At high drive amplitudes, these resonators exhibit nonlinear behavior, useful for studying fundamental effects such as stochastic resonance \cite{Badzey2005}, parametric amplification and frequency entrainment \cite{Shim2007}, and logic operation \cite{Guerra2010}. The nonlinearly-driven resonator can be further modulated with a signal at a separate modulation frequency. In the frequency domain, the response of the nonlinear resonator to the modulation appears as a sideband to the original resonator frequency that carries all the information encoded into the modulation signal. By decoding the sideband signal, the original signal can be retrieved.

Radiation pressure has been used in many applications including cooling micromirrors \cite{Gigan2006,Arcizet2006}, affecting optomechanical dynamics in cavities \cite{Carmon2005,Kippenberg2008}, and has been observed as a form of shot noise \cite{Purdy2013}. Existing devices and previous experiments have used optical radiation to perform line-of-sight communication \cite{Chan2000}, typically using photodiodes \cite{Brandl2014,Peyronel2016},  as the receiver. While these devices can be small and have the ability to communicate at a high bit rate, they can be highly wavelength dependent \cite{Huang2006,Carey2005},  and saturated by the presence of ambient light.

Several long-standing challenges with standard radio-wave wireless transmission still exist, including limited spectrum \cite{radiospectrum}; the last-mile problem \cite{cordeiro2003last} or infrastructure difficulty in the final leg of communication in the network to deliver services to end users; lack of indoor positioning system for accurate location determination \cite{vasisht2016decimeter}; and keeping secure transmission of data from potential compromise. Here, inspired by the first-ever wireless transmission by Alexander Graham Bell \cite{bell1880production} in 1880, we demonstrate that the force generated by the radiation pressure of an optical beam \cite{maxwell1881treatise} can be used as the modulation force added to the driven nonlinear resonator. Because our sideband modulation technique requires a non-zero modulation frequency, it is not affected by ambient conditions such as continuous solar radiation. The mechanical nature of this method has the added benefit that it is dependent on the intensity of the carrier and not its wavelength.

\section*{Materials and Methods}
As the intensity of the laser light modulates according to an encoded signal at a frequency $f_m$, the force adding to the nonlinear resonator being driven at a fixed resonator frequency $f_r$ generates the side band. The resonator can be described as a driven nonlinear oscillator with equation of motion \cite{Imboden2014}.
\begin{equation}
m \ddot{x} + \gamma \dot{x} + k x + k_3 x^3 = A_r \cos \left( 2\pi f_r t\right) + A_m S(f_m t),
\label{eq:eom}
\end{equation}
Here, $S(f_m t)$ represents the signal being transmitted by the laser, $x$ is the degree of freedom, $m$ is the effective modal mass, $\gamma$ is the damping coefficient, $k$ is the effective modal stiffness, $k_3$ is the cubic nonlinear spring constant, $A_r$ is the drive amplitude, and $A_m$ is the laser drive coefficient. The laser drive coefficient is proportional to the force exerted by radiation pressure, $F_{rad}=(R+1)I_{laser}A/c$, where $R$ is the surface reflectance, $I_{laser}$ is the beam intensity, $A$ is the resonator's effective mode area, and $c$ is the speed of light. The steady-state response was numerically shown to approximately obey the relation
\begin{align}
x(t) \approx c_r &\cos(2\pi f_r t) + c_m \cos(2\pi f_m t) \nonumber \\
&+ \sum\limits_n c_3(n) \cos\left(2\pi (f_r \pm n f_m) t \right)
\end{align}
where $n$ is an integer. The constants $c_r$, $c_m$, and $c_3(n)$ represent the sizes of the frequency responses. To verify Equation (2), we numerically solved Equation (1) and performed a fast Fourier transform on the results to obtain the frequency spectrum. Results are in the Supplemental Information.

A schematic diagram of our setup is shown in Figure 1(a). A green laser diode that is normally incident on the driven resonator provides the additional small signal using a pulse train with a modulation frequency between 1 and 5 kHz. The laser is modulated in amplitude and frequency using a waveform generator which was connected to the modulation port of the laser's driver. Several different waveforms, including sine waves and pulse trains, were tested, and all produced similar results. The MEMS resonator, shown in Figure 1(b), is a 96-by-270-$\mu$m rectangular plate resonator that is suspended by sixteen 15-by-3-$\mu$m legs. From bottom to top layer, it is constructed from 5-$\mu$m thick silicon and 1-$\mu$m thick silicon dioxide structural layers, a 300-nm thick molybdenum layer used for grounding, a 1-$\mu$m thick aluminum nitride (AlN) piezoelectric layer for signal transduction, and 300-nm thick interdigitated molybdenum electrodes for actuation and detection. Direct actuation and detection can be performed via a direct electrical connection to the marked electrodes in Figure 1(b). The other four electrodes, marked ``G", are used as ground references. The resonators are directly actuated via the inverse piezoelectric effect, and the response is measured using the direct piezoelectric effect. Applying a voltage to one of the transduction electrodes at some frequency causes a time-varying stress within the resonator. Likewise, stresses produce electric fields within the AlN which can be detected at the other transduction electrode. The device is symmetric, so either signal electrode can be used for either purpose.

\begin{figure*}
\includegraphics[width = 0.8\textwidth]{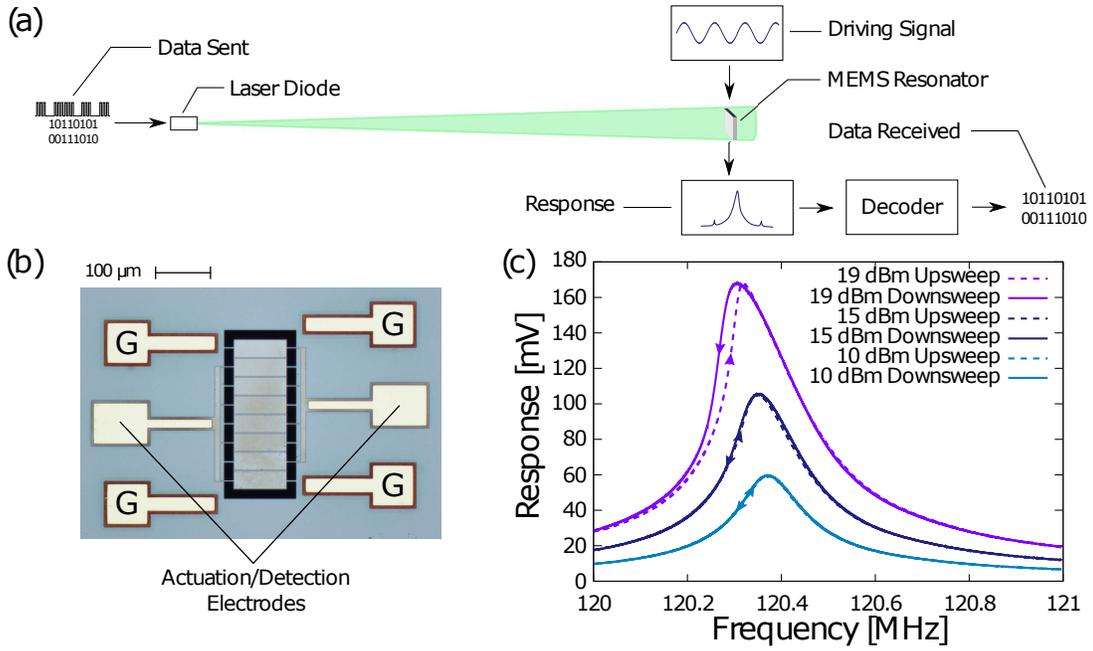}
\caption{(a) Schematic diagram of experimental setup used for data transfer. A 520-nm laser diode (LD) is modulated by a pulse train produced by an waveform generator and transmitted to the LD driver. The LD's beam is normally incident on a MEMS resonator, which is in a vacuum chamber at a pressure near $10^{-3}$ Torr. An excitation voltage is sent to the resonator using a signal generator and the response of the resonator is measured using a spectrum analyzer. The specific equipment used is listed in the Supplemental Information. (b) Top view of the MEMS resonator. The resonator is 96 $\mu$m $\times$ 270 $\mu$m and suspended using sixteen 3 $\mu$m $\times$ 15 $\mu$m legs. From bottom to top, the resonator is constructed from 5-$\mu$m silicon, 1-$\mu$m silicon dioxide, 300-nm molybdenum, 1-$\mu$m aluminum nitride, and 300-nm molybdenum. The silicon and silicon dioxide layers are the primary structure, the molybdenum layers are ground and actuation electrodes, and the aluminum nitride layer is used for signal transduction. The electrodes marked ``G" are used for grounding. (c) Frequency spectrum of MEMS resonator for 10-, 15-, and 19-dBm excitations. The low-power excitations produce Lorentzian responses, while the high-power spectra skew toward lower frequencies and show hysteretic behavior.}
\label{fig:fig1}
\end{figure*}

\section*{Results}
To characterize the resonator, the spectrum was measured (with the laser turned off) by sweeping the signal generator in the frequency range 120 to 121 MHz for a range of powers. The response of the resonator is measured using a swept spectrum analyzer. A subset of these measurements is shown in Figure 1(c). Using COMSOL, we numerically identified the mode to be a high-order Lamb wave mode with its largest deflections at the locations of each of the electrode pads (details in the Supplementary Information). At low powers, the resonator has a linear, Lorentzian frequency response. With increasing power, the response becomes pronouncedly nonlinear, which is evident from the skewing of the resonance peak and the onset of hysteretic behavior. This resonator's most sensitive resonance is at 120.4 MHz and has a quality factor of 485. For the remainder of the experiment, we directly drive the resonator at this frequency using a power of 19 dBm, providing a 10-$\mu$N effective force as estimated using a first-order calculation and verified numerically.

Next, the laser diode, with wavelength 520 nm, was turned on. Using the optical beam and resonator properties, we estimate the radiation force to be near 0.72 pN, orders of magnitude smaller than the resonator's driving signal (calculation details are in the Supplemental Information). With the laser modulation added, the presence of equally-spaced sidebands was observed, as shown in Figure 2(a). The first order upper-sideband, located at frequency $f_r+f_m$, was then characterized by driving the resonator in the range 120 to 121 MHz and measuring the sideband amplitude. The laser was modulated with a 50\%-duty-cycle pulse train at frequency 3 kHz. Figure 2(b) shows the size of the sideband and the downsweep response of the resonator when driven at 19 dBm. As expected, the sideband amplitude is related to the size of the resonator's peak response with the largest sideband amplitude appearing when the driving frequency is near the peak response frequency of the resonator. Further device characterizations are in the Supplementary Information.

\begin{figure*}
\includegraphics[width = 0.8\textwidth]{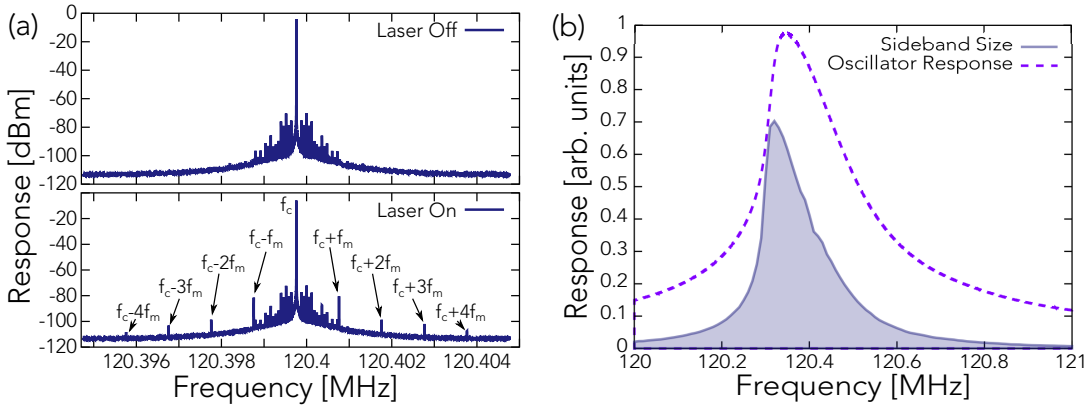}
\caption{(a) The top plot shows the spectrum of the resonator when driven at a single frequency by the signal generator, but no laser is incident. The bottom plot shows the spectrum of the same resonator when driven by the signal generator and a 1 kHz-modulated laser is also incident on the resonator. With the laser incident, equally spaced sidebands appear symmetrically about the resonator's driving frequency. The additional peaks near the central frequency in both plots are a result of sidebands produced by 60 Hz interference and by other nearby systems operating at low vibrational or electronic frequencies. (b) The shaded blue region of this plot shows the size of the upper sideband as a function of resonator excitation frequency for a laser modulation frequency of 3 kHz. The dashed purple line shows the same 19-dBm downsweep frequency spectrum as in Figure 1c.}
\label{fig:fig2}
\end{figure*}

Using these sidebands, wireless data transmission can be achieved using several different methods. First, we transmitted using Amplitude Modulation (AM) at a single modulation frequency using a 50\%-duty-cycle pulse train. To transmit a ``1", the pulse train has a large amplitude; to transmit a ``0", it has a smaller or zero amplitude. Figure 3(a) shows a portion of the AM transmission of the string ``Hello world". The purple line represents the logic levels that were transmitted, and the blue line indicates the amplitude of the measured sideband. Both data sets are normalized such that the minimum measurement is 0 and the maximum is 1. In this case, the ``1's" are transmitted using a 119.6 mA peak current to the laser diode and the ``0's" are transmitted by providing no current.

\begin{figure*}
\includegraphics[width = 0.8\textwidth]{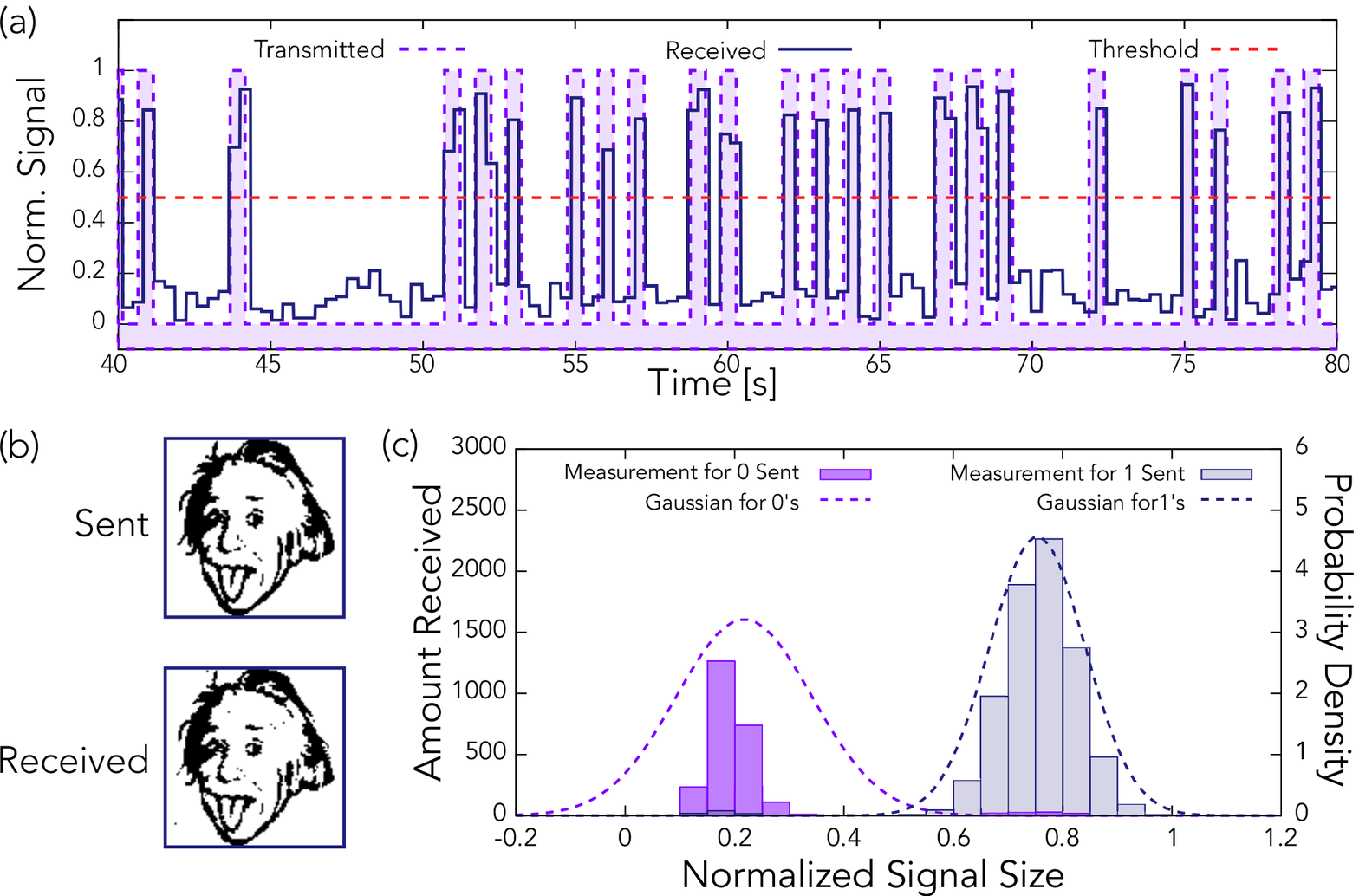}
\caption{(a) A selection of binary data that was transferred using an amplitude-modulated laser signal. The shaded purple region shows the transmitted data, the blue line shows the normalized sideband size (received data), and the dashed orange line shows the threshold between a 0 and 1 logic level. Each transmitted bit is separated by a ``0". Signals are normalized such that the minimum possible signal is 0 and the maximum possible signal is 1. The sideband is measured by the spectrum analyzer using an 18 Hz bandwidth, and is recorded using SCPI commands to locate and record the peak. (b) The image of Einstein in the ``Sent" box was transmitted to the MEMS resonator using the amplitude-modulated laser. The image in the ``Received" box represents the data that was received using the resonator. Each image is constructed from 10,000 pixels. Of the 10,000 pixels transmitted, 155 were misinterpreted. (c) The bars show the distribution of normalized signal sizes that were actually received during the AM transmission of the Einstein image. Purple bars are signal sizes measured when a ``0" was transmitted, and blue bars were measured when a ``1" was transmitted. The dashed lines are normalized Gaussian distributions based on the received data. The poor fit of the ``0" Gaussian is a result of the large number of errors that resulted in reading of a ``0" transmission as a ``1".}
\label{fig:fig3}
\end{figure*}

After ``Hello world!" was successfully transmitted several times, we transmitted a larger file -- a 100-by-100 pixel monochrome version of the iconic image of Einstein shown in the top panel of Figure 3(b). White pixels were represented by ``1's" and transmitted using a 140-mA pulse train; black pixels were represented by ``0's" and transmitted using a 59.6-mA pulse train. The results of this transmission, which contained 155 errors, are shown in Figure 3(b). We performed a statistical analysis of the data received and found that the amplitudes of ``0" and ``1" data points obey Gaussian distributions with standard deviations 0.12 and 0.087, respectively, as illustrated in Figure 3(c). From these distributions, we obtain an average Bit Error Rate (BER) of 1 error per 86.9 bits.

An alternate method of transmitting data used was Frequency Modulation (FM). We again used a 50\%-duty-cycle pulse train that varied between a maximum current of 140 mA and a minimum current of 0 mA; however, in this case, the pulse train was not turned off at any point during transmission. Instead, data was transmitted by making minor adjustments to the frequency of the pulses in order to change the frequency at which the sideband occurs. By assigning each side band frequency a ``value", non-binary digital data was transmitted.

In Figure 4(a) we present a hexadecimal data transmission. For the data shown, laser modulation of 1 kHz ($f_0$) represents a ``0" and successive digits are spaced in 50 Hz ($\Delta f$) increments. A ``-1" (for our parameters, laser modulation at 950 Hz) was transmitted between digits to aid in processing the received data. The largest frequency measured between two ``-1's" was stored, and the function $digit=(f_{stored}-f_0)/\Delta f$  was rounded to the nearest integer to decode the signal. Figure 4(b) shows the result of a transmission of the same image of Einstein as before which, unlike our AM transmission, contains no errors. For this transmission, groups of four bits were encoded into a single hexadecimal digit.

\begin{figure*}
\includegraphics[width = 0.8\textwidth]{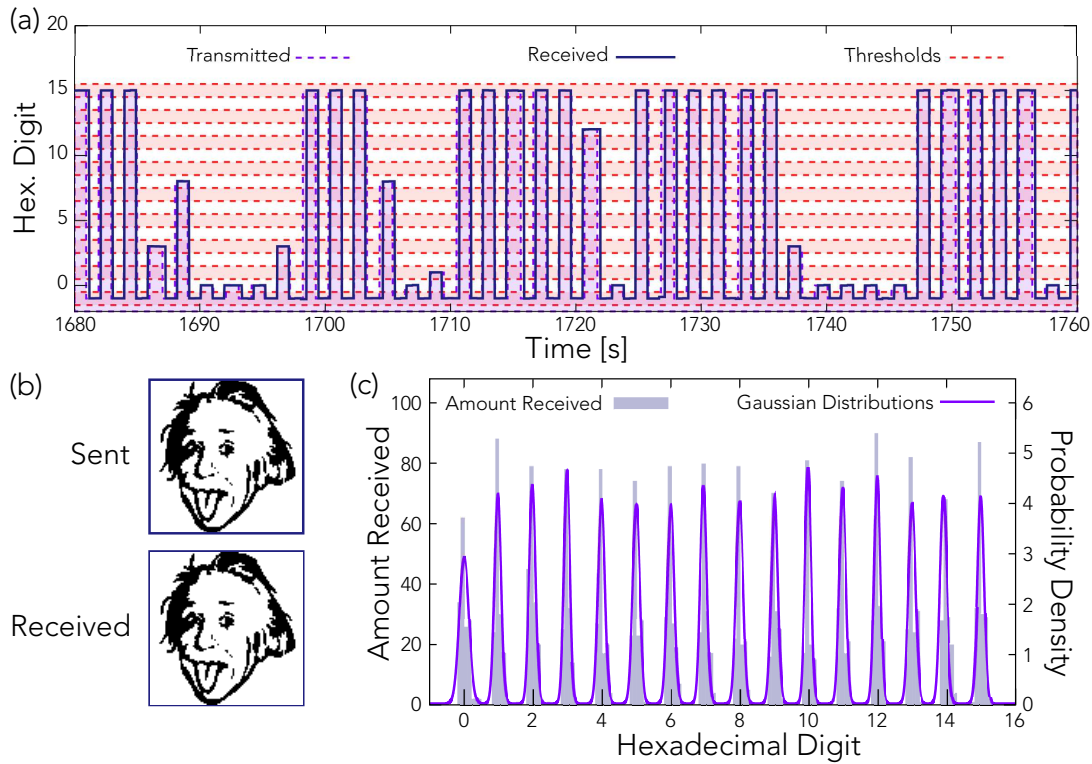}
\caption{(a) This plot shows a selection of hexadecimal data that was transferred using a frequency-modulated laser signal. A continuous signal with time-varying frequency was used to transmit this data. The shaded purple region shows the transmitted data, the blue line shows the interpreted hexadecimal digit as calculated using $(f_{sideband}-f_0)/\Delta f$ without rounding, and the dashed orange lines show the thresholds between hexadecimal values. Transmitted digits are separated by a ``-1" signals. The measurement bandwidth is 12 Hz for all FM data transmission results presented. (b) The image of Einstein in the ``Sent" box was transmitted to the MEMS resonator using the frequency-modulated laser. The image in the ``Received" box represents the data that was received using the resonator. Each image is constructed from 10,000 pixels. (c) The blue bars show the distribution of normalized signal sizes that were actually received during a randomized data transmission. The purple lines show the normalized Gaussian distributions based on the received data.}
\label{fig:fig4}
\end{figure*}

Next, we transmitted a series of random hexadecimal digits to estimate the BER. As shown in Figure 4(c), the distributions of sideband frequencies for each hexadecimal digit approximately obeyed Gaussian distributions with standard deviations averaging 5.6 Hz. For FM transmission, we estimate an average BER of 1 error per 56,320 data packets.

\section*{Discussion}
While the specific equipment used in our proof-of-concept demonstration is not suitable for practical applications, alternate arrangements including a combination of more powerful lasers, use optics designed to focus beam at resonator location, and arraying resonators may be effective in producing higher signal-to-noise ratios and bitrates. In addition, our demonstration was not designed for high speed applications, as its speed was severely limited by the speed of the spectrum analyzer. In principle, this method can be used for much higher bitrates than demonstrated here by using high-frequency, low-$Q$ resonators. Reducing $Q$ also has the benefit of increasing the usable bandwidth, but reduces the signal size, so the resonator must be carefully designed for its specific application. Some nanomechanical resonators have been shown to have frequencies on the order of 10 GHz \cite{Rinaldi2009}, so the maximum feasible bitrate, which is limited to approximately $f_0/Q$ in an open-loop configuration, can be hundreds of MHz to several GHz. Of course, the modulation frequency must be larger than the bitrate in order to consistently produce measurable sidebands.

There are additional practical limitations to this method, but many can be easily addressed. For one, distance, beam intensity, resonator angle relative to beam, and other factors can cause fluctuations in the intensity of the sideband which can corrupt data transmitted via amplitude modulation. We experienced this behavior during our experiment as well; sideband amplitudes fluctuate somewhat randomly, but their frequencies are highly predictable. For this reason and because of the much more favorable BER, transmission should be performed using FM. Optimal sideband frequencies should also be determined prior to implementation; low frequencies are susceptible to noise from sources such as ambient vibration and electrical signals. Sideband frequencies should also be chosen to avoid noise from known sources of mechanical vibrations.

In addition to successful demonstration of wireless data transmission, we performed control experiments to verify that the sideband response and its dependence on the modulation signal is indeed due to the modulation in the radiation pressure exerted on the resonator. To this end, blocking the resonator from the laser light produced no sideband signal. Further, we found that the sideband size is proportional to the optical power of the laser as predicted for excitation by radiation pressure. To verify that the resonator was not being actuated by photothermal effects \cite{Lammerink,Evans2014}, temperature change due to radiation was monitored using the shift in the resonance frequency. As discussed in Supplementary Information, though the resonance frequency of the resonator shifts with heating, it does not produce a sideband even when heating is modulated in a similar manner as the laser. While the resonance frequency changes with temperature, the frequency of the sideband is related to the carrier frequency, which is externally generated and not a function of temperature. The main effect of temperature changes is that the shifting of the resonance can affect the amplitude of the sideband. Since the sideband amplitude is related to the resonator's response at the carrier frequency, a feedback loop should be implemented in practice to adjust the resonator driving frequency with a measured temperature. To further verify that the sideband is generated due to nonlinear mixing, the resonator was driven in the linear regime, which produced no sideband, reconfirming Figure 2.  

\section*{Conclusion}
In conclusion, we demonstrated successful wireless transmission of data using the force produced by a phenomenon as weak as radiation pressure. Depending on the modulation, this method can be used to transmit digital or analog information. Applying techniques in signal demodulation and arraying resonators to increase the signal size and fidelity, our methods can be adapted for use in technologies ranging from signal transduction in fiber optic communications systems to virtually undetectable free-space communications with satellites, as can be done with similar techologies \cite{radiospectrum,cordeiro2003last,vasisht2016decimeter}. Our method has the added advantage being a purely mechanical effect and hence is not strongly wavelength-dependent as photodiodes are known to be.

\bibliography{references}

\end{document}